\documentclass[pra,showpacs,12pt]{revtex4}

\usepackage{epsfig}
\usepackage{amsfonts}

\begin{document}

\title{Two-spin subsystem entanglement in spin $1/2$ rings with long range interactions}
\author{M. Gaudiano}
\email{gaudiano@mate.uncor.edu}
\author{O. Osenda}
\email{osenda@famaf.unc.edu.ar}
\author{G.A. Raggio}
\email{raggio@famaf.unc.edu.ar}
\affiliation{FaMAF-UNC, C\'ordoba, Argentina}
\begin{abstract} We consider the two-spin subsystem entanglement for eigenstates of the Hamiltonian
\[ H= \sum_{1\leq j< k \leq N} \left( \frac{1}{r_{j,k}}\right)^{\alpha} {\mathbf \sigma}_j\cdot {\mathbf \sigma}_k \]
for a ring of $N$ spins $1/2$ with asssociated spin vector operator $(\hbar /2){\bf \sigma}_j$ for the $j$-th spin. Here $r_{j,k}$ is the chord-distance betwen sites $j$ and $k$.\\
The case $\alpha =2$ corresponds to the  solvable Haldane-Shastry model whose spectrum has very high degeneracies not present for $\alpha \neq 2$. Two spin subsystem entanglement shows high sensistivity and distinguishes $\alpha =2$ from $\alpha \neq 2$. There is no entanglement beyond nearest neighbors for all eigenstates when $\alpha =2$. Whereas for $\alpha \neq 2$ one has selective entanglement at any distance for eigenstates of sufficiently high energy in a certain interval of $\alpha$ which depends on the energy.\\
The ground state (which is a singlet only for even $N$) does not have entanglement beyond nearest neighbors, and the nearest neighbor entanglement is virtually independent of the range of the interaction controlled by $\alpha$. 
\end{abstract}
\date{31/08/07}

\maketitle

\section{Introduction }

Since the first studies  \cite{orig1,orig2,orig3} of  entanglement in the ground-state of interacting spin $1/2$ systems,
 a considerable amount of work has been devoted to analyze this feature. The Hamiltonians most studied have been those with nearest neighbor interactions ($XX$, $XY$, $XYZ$, $XXZ$  etc.) between the spins in the presence of an external magnetic field (usually constant) which is the order parameter for a quantum phase transition (i.e., non-analyticity in the ground- state energy; \cite{Sa}).  The fascinating and intricate  connections and relations between (mainly two-site subsystem) entanglement and the quantum phase transition have been systematically studied after \cite{OsAmFaFa,OsNi}.
At the other extreme, namely models where each spin interacts identically with all the others, there are studies of the Lipkin-Meshkov-Glick model (\cite{DuVi1}) which is of the $XY$ type in an external field, and of an equivalent of the BCS model (\cite{DuVi2}) which is of the $XX$ type in an external field. A recent review of the subject is \cite{Faetal}.\\
Here, we are not concerned with the quantum phase transition aspects but with the dependence on the range of the spin pair-interactions.  One of the basic facts emerging from the studies mentioned above, is that if the anisotropies are large in nearest-neighbour interacting systems, then the system behaves qualitatively as the (quantum) Ising model ($\sum_{j} \sigma_j^x\sigma_{j+1}^x + h \sigma_j^z$). Also, entanglement vanishes for large external fields. Moreover, two-site subsystem entanglement in the ground-state is short-ranged: it vanishes if the sites are not nearest, or next-nearest neighbors.
\\

The maximization of nearest-neighbor entanglement in translationally invariant pure states of quantum  spin $1/2$ chains has been the subject of
various studies. \cite{orig3} finds an upper bound of about $0.434$ for the nearest neighbor concurrence, and shows
that in  antiferromagnetics rings  in which
neighboring particles interact via the Heisenberg Hamiltonian there are states
with larger nearest-neighbor entanglement than the ground state.  On the
other hand, since the work of Coffman, Kundu and Wootters, \cite{coffman2000}, it is
known that there are limitations to the amount of entanglement that can be shared by 
three q-bits, so some efforts have been made in order to find if in a uniform
chain where each qubit is equally entangled with its two
neighbors, the Coffman-Kundu-Wootters bound is achievable. The subject was also addressed  in \cite{wootters2002}. \\
Nearest-neighbor entanglement was also analyzed  in \cite{benatti2005,hiesmayr2006,MaNa}, for the finitely correlated states (generalized valence bond states) of Fannes, Nachtergaele and Werner \cite{FaNaWe}, which arise as ground states of Hamiltonians with short-ranged interactions. The upper bound for nearest-neighbor concurrence obtained in \cite{hiesmayr2006} is practically equal to that of \cite{orig3}.

 In this communication we present some results on entanglement of 2-spin subsystems for eigenstates of a Hamiltonian where the spins are subject to a long-range interaction inversely proportional to a power of their distance and the external magnetic field vanishes. Our study is not restricted to the ground-state(s), but includes the whole spectrum.  In particular, we are interested in which
eigenstates show two-site entanglement for long
distances, i.e. beyond nearest or next-nearest neighbors. Besides, we are
interested in the dependence of the distance for  which it is
possible to obtain two-site  entanglement on the range of the interaction
between the spins.

\section{The model} Specifically we consider $N$ spins $1/2$ and the Heisenberg Hamiltonian is given by
\[ H_N ( \alpha ) = \sum_{1\leq j< k \leq N} \frac{ {\mathbf \sigma}_j\cdot {\mathbf \sigma}_k}{( r_{j,k})^{\alpha}}\;;\]
where ${\mathbf \sigma}_j=(\sigma_j^x,\sigma_j^y,\sigma_j^z)$ is the vector operator formed with the three Pauli operators associated to the $j$-th spin, and 
\[ r_{j,k} = \frac{ \sin (\pi |j-k|/N) }{\sin (\pi / N)} \;,\;\; j,k=1,2,\cdots , N\;,\]
is proportional to the distance between vertexes $j$ and $k$ in a regular, flat, $N$-gon whose vertices are numbered consecutively. The constant in the definition is chosen so that the nearest neighbor distance $r_{j,j+1}$ is one. For given $N$, the number of distinct distances is $[N/2]$ --the largest integer not above $N/2$. We observe that $H_N(0)$ is the isotropic Heisenberg model where each spin interacts identically with every other spin.  For $N\geq 2$ fixed,  the limit $\alpha \to \infty $ corresponds to the nearest-neighbor antiferromagnetic XXX model.  We often drop the index $N$ and parameter $\alpha$ in $H_N(\alpha )$, when these are irrelevant.\\

The Hamiltonian $H$ is ``anti-ferromagnetic'', while $-H$ is  ``ferromagnetic''.
If $N$ is even then the least energy eigenvalue is non-degenerate, whereas it is degenerate for uneven $N$. \\
For $\alpha =2$ this is the Haldane-Shastry model \cite{Hal,Sha} which is explicitely solvable \cite{yang,french}. Here the least eigen-energy is fourfold degenerate for uneven $N$, and the largest energy eigenvalue (ground-state energy of the ferromagnetic version) is always degenerate with multiplicity $N+1$.\\ 
The spectrum for $\alpha =2$ is highly degenerate with comparatively few eigenvalues and very different from the spectrum for $\alpha \neq 2$ and same $N$. For example for $N=8$, where the Hilbert space has dimension $2^8=256$, $H$ has 5 eigenvalues for $\alpha =0$, 19 for $\alpha =2$ and 45 for $0< \alpha \neq 2$, except for a discrete and finite set of values of $\alpha$, where a crossing or two reduces the number of distinct eigenvalues by 1 or 2 (see figures \ref{Evsalfa}, \ref{sizeofspectrum}); the nearest-neighbor XXX model corresponding to $\alpha = \infty$ has $40$ eigenvalues. These qualitative features do not depend on $N$.

Due to the absence of effective criteria for multipartite entanglement in mixed states, we study only the entanglement of the possible pairs of spins, that is two-site entanglement. It suffices to consider the pairs $(1,j)$ for $j=2, \cdots , [N/2]+1$; corresponding to the possible distances. Given a state $\rho$ of system of $N$ spins, $\rho_{j,k}$ denotes the reduction of $\rho$ to the subsystem with components (i.e., sites) $j$ and $k$. The entanglement of this reduced state is detected and quantified by its concurrence, \cite{Wo}.\\

Our analysis proceeds as follows. One has
\[ {\mathbf \sigma}_j\cdot {\mathbf \sigma}_k = 2\Pi_{j,k}- {\bf 1}\;,\]
where $\Pi_{j,k}$ is the transposition interchanging the $j$-th and $k$-th components of the product basis vectors
\begin{equation}
 |\epsilon_1, \epsilon_2, \cdots , \epsilon_N \rangle\;,\;\; \epsilon_n = \pm\;,\;\; n=1,2,\cdots , N \;,\label{base} 
\end{equation}
where $\sigma^z |\pm\rangle = \pm \,| \pm \rangle$. That is,
\[ \Pi_{j,k} |\epsilon_1, \epsilon_2, \cdots , \epsilon_j , \cdots , \epsilon_k , \cdots ,\epsilon_N \rangle =
 |\epsilon_1, \epsilon_2, \cdots , \epsilon_k , \cdots , \epsilon_j , \cdots ,\epsilon_N \rangle\;.
\]

To simplify the structure of the corresponding matrix in the above basis, we consider the operator (Hamiltonian)
\[ \widetilde{H}_N( \alpha )= \frac{1}{2}\sum_{1\leq j< k \leq N} \left( \frac{1}{r_{j,k}}\right)^{\alpha} (\Pi_{j,k}- {\mathbf 1})  \]
which differs from $H_N ( \alpha )$ by an additive, $(N, \alpha )$-dependent constant. Thus $H_N ( \alpha )$ and $\widetilde{H}_N( \alpha)$ have the same number of distinct eigenvalues with the same multiplicities and the same spectral orthoprojectors. Figure \ref{Evsalfa} shows the spectrum of $\widetilde{H}$ for $N=8$ as a function of $\alpha$ up  to $\alpha =5$. We find that beyond $\alpha$ about $7.29$ there are no crossings, c.f. figure \ref{sizeofspectrum}, but eigenvalue curves do approach each other asymptotically leading to $40$ eigen-energies in the nearest-neighbor model ($\alpha = \infty$).  The largest eigen-energy of $\widetilde{H}_N( \alpha )$ is zero, it is  $(N+1)$-fold degenerate for every $\alpha \geq 0$, and the corresponding spectral orthoprojector is independent of $\alpha$ \cite{Gau}. In fact an orthonormal basis of this eigen-space is easily described; is consists of $N+1$ vectors each of which is the normalized sum of the $\left( \begin{array}{c} N\\ s\end{array}\right)$ vectors of the form (\ref{base}) where exactly $s$ of the $\epsilon_n$'s are $+$; and $s=0,1, \cdots , N$.\\
 
\section{Entanglement}

Denote the spectrum with $spec$; suppose
\[ H = \sum_{E \in spec(H)} E P_E \;,\;\; P_EP_{E'}=\delta_{E,E'}P_E \;,\;\; \sum_{E\in spec(H)}P_E = {\mathbf 1} \;,\]
is the spectral decomposition of the Hamiltonian $H$. Then $tr(P_E)$ is the multiplicity (degeneracy) of the eigen-energy $E$. \\

At this point, we must recall Theorem 2 of \cite{SaTaVi} which says that any subspace of dimension at least $2$ of the four-dimensional Hilbert space of two q-bits contains at least one product vector. Thus if the eigen-energy $E$ of a two  spin $1/2$ system is degenerate, then there is a separable eigenvector to $E$. Extensions of this result to sufficiently high dimensional subspaces of $N$ ($>2$) q-bits are possible but this is not the point of this paper.  To analyze entanglement in the case of degenerate eigenvalues, we must consider the (uniform) eigen-state
\[ \rho (E):= P_E/tr(P_E) \;,\;\; E \in spec (H)\;.\]
Notice that this state is obtained by mixing with equal weights (namely $1/tr(P_E)$) any (pairwise orthogonal) pure eigen-states whose corresponding vectors constitute an orthonormal basis of the eigenspace of the eigenvalue $E$ of  $H$.\\

For the Haldane-Shastry model ($\alpha =2$), we use the known eigenvalues and degeneracies \cite{french}, and determine the orthoprojectors by: $  P_E = \prod_{E'\neq E} \frac{H-E'}{E-E'}$. 
For $\alpha \neq 2$ we determine the eigen-energies and corresponding spectral projections numerically.
Due to the particular symmetry of $H$ or $\widetilde{H}$ which commutes with  $\sum_{j=1}^N \sigma_j^z$ and with $\prod_j \sigma_j^z$ (\cite{Sul,Gau}) the reduced density operators (in the basis (\ref{base}) of product eigenvectors of $\sigma_j^z$, $j=1,2,\cdots , N$) for any pair $(j,k)$ of spins have the same structure namely
\[ \left( \begin{array}{cccc} a&0&0&0\\
0&b&c&0\\
0&c&b&0\\
0&0&0&a \end{array}\right) \;,\]
where $ a,b \geq 0$, with $a+b=1/2$; and   $ c$ is real with $|c|\leq b$.
The concurrence of this 2-spin state is $\max \{ 0 , 2 \max \{ a, b+|c|\}-1\}= \max \{ 2(|c|-a),0\}$.\\

The above structure of $\rho_{j,k}$ implies that the reduced density matrix for the $j$-th site is simply and always just $(1/2)I$ which is the maximally mixed state for a spin $1/2$. This remarkable feature of the model is independent of $N$, $\alpha >0$ and the eigen-energy considered.
As a consequence, the Meyer-Wallach measure \cite{MeWa} (which, when the state is pure, is a true measure of entanglement and not just a measure of degree of mixture) given by $2 - (2/N) \sum_{j=1}^N tr ( \rho_j^2)$  is identically equal to $1$ for every eigen-state $\rho (E)$ independently of $N$, $\alpha >0$ and the eigen-energy considered.\\

Another remarkable feature is that $\rho_{j,k}(0)$, the reduced density matrix for sites $(j,k)$ for the eigen-state of maximal energy which was described above and seen to be independent of $\alpha$, is also independent of the pair $(j,k)$, \cite{Gau}. It turns out that the this maximal energy eigen-state (ground-state of the ferromagnetic model) does not show entanglement at  all site distances.\\

Our calculations were performed for $N=2,3,4,5,6,7,8$. Up to the degeneracies in the lowest eigen-energy, the qualitative features are independent of $N$ in that range. The graphs shown corrrespond to $N=8$.\\

Here is a list of some of our observations for the mentioned values of $N$ (others will follow):
\begin{itemize}
 \item The isotropic model $H_N(0)$ shows no two-site entanglement for all distances at every eigen-energy. 
 \item In the ground-state there is exclusively nearest neighbor two-site entanglement  for every $\alpha >0$ ($\alpha >0$ was sampled rather completely only for $N=8$); the corresponding concurrence is a slowly varying increasing function of $\alpha >0$ which is discontinuous at $0$ (see figure \ref{fig12-13}).
\item For every $\alpha >0$ nearest neighbor entanglement appears only in the first few energy levels (i.e., for $N=8$, the first four or five energy levels). If an excited state presents nearest neighbor entanglement for some $\alpha >0$ then the corresponding concurrence is below that of the ground-state nearest neighbor concurrence for that value of $\alpha$. However, nearest neighbor concurrence is not generally a decreasing function of the energy for fixed $\alpha >0$ (see figure \ref{a=1}).
\item The distinctive feature of the case $\alpha =2$ with respect to the cases $\alpha\neq 2$ is simple: {\em For $\alpha =2$ there is no two-site entanglement beyond nearest neighbors at all eigen-energies. For $0 < \alpha \neq 2$ two-site entanglement for other possible distances appear at some excited eigen-energy level.} In fact, two-site entanglement for all possible distances is present for every $\alpha$ at some excited energy-level  except for $\alpha$ in a certain bounded interval  which depends on $N$.

\item Except for exceptional values of $\alpha$ one finds excited states where the concurrence for sites further apart than nearest neighbors have greater concurrence than the ground-state nearest neighbor concurrence. For example, for $N=8$ and almost all $\alpha >0$ ($\alpha =1$ is exceptional) one finds some excited state where the concurrence for sites at maximal distance  is always above the nearest neighbor concurrence for the ground state.

\item In the ferromagnetic model $-H$, the ground-state does not show pair-entanglement at all distances and the same is true for the low energy eigenstates, but the number of the states with this property decreases with $\alpha$. The first excited states which exhibit two-site entanglement do so for the largest distances. Only high energy eigen-states show nearest-neighbor entanglement. All this is seen in figures \ref{a=1}-\ref{a=nn}. \end{itemize}
 
The series of four figures \ref{a=1}-\ref{a=nn} show the concurrence of $\rho_{j,k} (E)$ for the possible eigen-energies $E$ for $N=8$ and distinct values of $\alpha$. In all these figures the dot corresponds to nearest neighbor distance ($|j-k|=1$), the cross $\times$ to the next-nearest neighbor distance ($|j-k|=2$), the star $\ast$ to $|j-k|=3$ and the empty square $\square$ to the maximal distance $|j-k|=4$. \\

We mention a curious feature of the nearest neighbor model which is apparent in figure \ref{a=nn}. Nearest neighbor concurrence where it is positive, is a linear decreasing function of the energy: Concurrence ( $\rho_{1,2}(E) ) = \max \{ -A E - B , 0 \}$, where $A,B >0$. This linear regime is reached very rapidly as $\alpha$ grows to infinity; e.g. for $N=8$ and $\alpha > 6$ one is practically in the linear regime.\\

For $N=8$ we have also analyzed if energy is distance selective for two-site entanglement. We find that if at some eigen-energy there is two-site entanglement at nearest or next-nearest neighbor distance then there is no entanglement at all the other possible distances. However, two-site entanglement at maximal and next to maximal distances can be present for the same energy level in certain intervals of $\alpha$ values.

Another way of presenting our results consists in graphing the concurrence of $\rho_{j,k}(E)$ for fixed $N$ and distance $|j-k|$ as a function of $\alpha$ while keeping the number of the (excited) level $E$ fixed. This is done in figures \ref{15-2}, \ref{14-8}, \ref{15&14-42}, and \ref{detail}. Figure \ref{15-2} --always for $N=8$-- gives the concurrence of the second excited eigenstate $\rho (E)$ (associated to the second eigen-energy above the ground-state energy) reduced to two sites at the maximal possible distance for $N=8$ which corresponds to $|j-k|=4$; after $\alpha \approx 4$, there is persistent entanglement for maximally distant sites in this energy eigen-state.\\
Figure \ref{14-8} illustrates the same features for the eighth excited state  and distance $|j-k|=2$ (next nearest neighbors); entanglement is only present in a bounded interval $(a,2)$ with $a$ below $1$. Figure \ref{15&14-42} presents two-site concurrence at maximal distance and distance $|j-k|=3$ for the fourty first excited energy eigen-state (which is not present for $\alpha =2$); again entanglement persists and coexists for both distances for this level beyond $\alpha \approx 7.29$. Below this value both concurrences show discontinuities as illustrated in figure \ref{detail} for the concurrence of sites at maximal distance in the fourty first excited state.\\

Figure \ref{cantidad}  shows the number of energy eigen-states with two-site entanglement at the four possible distances present in the case $N=8$ as a function of $\alpha$.\\

We also calculated  for completeness the measure of de Oliveira et al \cite{deOl}:
\[ E^{(2)}_{gl} = \frac{4}{3} \frac{1}{N-1} \sum_{j=1}^{N-1} \left( 1 - \frac{1}{N-1}\sum_{k=1}^N tr (\rho_{k,k+j}(E)^2)\right)\;,\]
for fixed $N$ and fixed number of the excited eigen-energy in $\rho (E)$ as a function of $\alpha$. 
We observe that this measure (which is a true multipartite entanglement measure only for pure states) is a slowly varying function of $\alpha$  except for jump discontinuities at all points  where the corresponding level experiences a crossing.  \\

The viewpoint adopted in the previous figures --holding the number of the excited energy level fixed-- which is often natural in physical problems, is not the most appropiate to obtain a simple description of our observations due to the large amount of crossings (apparent in figure \ref{Evsalfa}). It is more appropiate to adopt a perturbation theoretic point of view. 
For fixed $N$, the operator families $H_N ( \alpha )$ and $\widetilde{H}_N ( \alpha )$ are holomorphic in $\alpha \in {\mathbb C}$ in the sense of Kato \cite{Ka}. Thus, for example for $N=8$, and away from the exceptional points where crossings occur (see figure \ref{Evsalfa}), there are fourty five pairwise orthogonal projectors which depend real analytically on $\alpha$.
The two-site concurrence for any pair of sites obtained by normalizing these orthoprojectors to states will be real analytic functions of $\alpha$ away from the crossings.  This alternative way of analyzing two-site entanglement, shows that only very few of the orthoprojectors carry two-site entanglement as is illustrated in the following figure pairs. Figure  \ref{fig12-13} 
 shows all the non-zero concurrences for nearest neighbors (namely five)  for $N=8$ as functions of $\alpha$ in the top graph, while the lower graph 
shows which five of the energy curves of figure \ref{Evsalfa} give rise to these nearest neighbor concurrences. In  the concurrence graph of figure \ref{fig12-13}, only the curve corresponding to maximal nearest-neighbor concurrence ($\approx 0.41$) is associated to only one level, in this case the ground-state. All other concurrence curves in this graph mix (due to crossings) energy levels of different excitation number.  The discontinuities occur at crossings and not all of them are visible in the figure.  The following pair of graphs \ref{fig14-15} deals with next-nearest neighbors. The energy curves (lower graph) are only drawn where two-site entanglement (at the corresponding distance) is present and they are not drawn through the whole range of values of $\alpha$ (this is repeated for the other distances to be shown below). Observe that the entanglement at next-nearest neighbor distance $|j-k|=2$ has a gap: for $\alpha$ betweeen about $2.54$ and $3.71$ there is no next-nearest neighbor entanglement at all eigen-energies; this is the only distance which shows this feature.  The graph pairs of figures \ref{fig16-17} and  \ref{fig18-19}, repeat this for the distances $|j-k|=3$, and $4$ respectively. Close inspection of figures \ref{fig16-17} and \ref{fig18-19} show that the maximal concurrence (labeled {\em a} in figure \ref{fig18-19}) at maximal distance $|j-k|=4$ and the monotone increasing concurrence curve (labeled {\em a} in figure \ref{fig16-17}) for the distance $|j-k|=3$ originate in the same energy curve (i.e. its corresponding orthoprojector) namely {\em a} in the lower graphs of figures \ref{fig16-17} and \ref{fig18-19}. This feature appears also for nearest neighbor and maximal distance entanglement: the energy vs. $\alpha$ curves labeled {\em c} in figures \ref{fig12-13} and \ref{fig18-19} coincide; however entanglement at maximal distance appears only above $\alpha \approx 3.88$
A careful count shows that of the $45$ orthoprojectors only $12$ carry two site entanglement, of these $10$ do so for one distance only, only one  carries two-site entanglement at two distances simultaneously for all values of $\alpha >0$, and only one carries two-site entanglement at two distance simultaneously albeit above some threshold value of $\alpha$  (as just described).\\

The following table gives the dimension of the fourty five orthoprojectors appearing for $N=8$ and the number of times each dimension appears.

\begin{center}

\begin{tabular}{c||c|c|c|c|c|c|c|c|c|c|c|c|c|c|c}
 dimension of orthoprojector & 1&2&3&4&5&6&7&8&9&10&11&12&13&14&$\geq 15$\\
\hline
number of orthoprojectors & 6&4&6&0&6&11&1&0&1&7&0&0&0&3&0
\end{tabular}
\end{center}

All six one-dimensional orthoprojectors carry two-site entanglement and two of these do so for two distinct distances ($|j-k|=3,4$ for all $\alpha >0$ and $|j-k|=1,4$ for $\alpha > 3.88$).
\section{Concluding remarks}

We present a study of pair (or two-site) entanglement for the eigenstates of an $N$ spin $1/2$ model where the spins (sites) are equidistant in a circular ring and the pair-interaction is inversely proportional to an arbitrary positive power $\alpha$ of the site distance, and proportional to the scalar product of the magnetic moments of the spins. The model interpolates between a nearest neighbor interaction model (of type XXX) and a model where every spin interacts equally with all the others. The model is solvable for $\alpha = 2$ (Haldane-Shastry model, loc. cit.) where the spectrum is extremely degenerate, for every $N$, relatively to $\alpha \neq 2$ and same $N$. We do not include an external magnetic field in the Hamiltonian, and thus exclude a quantum phase transition, i.e. non-analytic behavior of the ground-state energy as a function of the magnetic field strength.\\
We do not concentrate on ground-state entanglement (the ground-state energy is degenerate for uneven $N$) but analyze all eigen-energies. The qualitative features reported are independent of $N$ up to $N=8$, and we have no doubts that they are independent of $N$ for all $N$.

One of our original motivations for studying the model was to find long-range pair entanglement in the ground-state! Our expectations in this direction were completely wrong;  a key feature  is  the observed insensitivity to the range of the interaction of two-site entanglement for the ground-state. Not only there is  no two-site entanglement beyond nearest neighbors independently of the range  of the interaction controlled by $\alpha$, but also, e,.g. for $N=8$, taking the limit $\alpha \to 0^+$ and comparing this with the  nearest-neighbor concurrence in the nearest neighbor interaction model ($\alpha = \infty$), the  variation is only about 2.5\% percent over the whole range of values of $\alpha >0$. Moreover the value is about $94 \%$ of the upper bound claimed by \cite{orig3}.\\

Our second observation is that a simple glance at a figure of the type of figures \ref{a=1}-\ref{a=nn} allows the onlooker to decide whether $\alpha=2$ or not. Two-site entanglement of eigenstates is able to detect a spectral ``collapse''.\\

Thirdly, one can produce selective two-site entanglement at any required distance by appropiately choosing $\alpha$ and/or the energy-level; a feature which is of some interest.\\

Finally, we point out that for every $2 \neq \alpha >0$, including the nearest neighbor model (see figure \ref{a=nn}), there is an excited eigen-state which is pure (except for special values of $\alpha$ where a crossing occurs; see figure \ref{fig18-19}) which presents entanglement at maximal distance and at next to maximal distance simultaneously. This state corresponds to the energy vs. $\alpha$ curve labeled {\em a} in figures \ref{fig18-19} and  \ref{fig16-17}. The concurrence for maximal distance in this state is the maximal concurrence for all possible distances in all eigen-states. It would be interesting to give an experimental procedure to prepare this state.

\newpage

$\;$

\begin{figure}
\begin{center}
 \psfig{figure=energies.eps,width=11.4cm}
\end{center}
\caption{\label{Evsalfa} Spectrum of $\widetilde{H}_8( \alpha )$. $0$ is the largest eigen-energy and has degeneracy $9$ ($=N+1$) for all $\alpha \geq 0$. Beyond $\alpha \approx 7.29$ there are no further crossings, c.f. figure \ref{sizeofspectrum}.}
\end{figure}

$\;$

\newpage
$\;$

\begin{figure}
\begin{center}
 \psfig{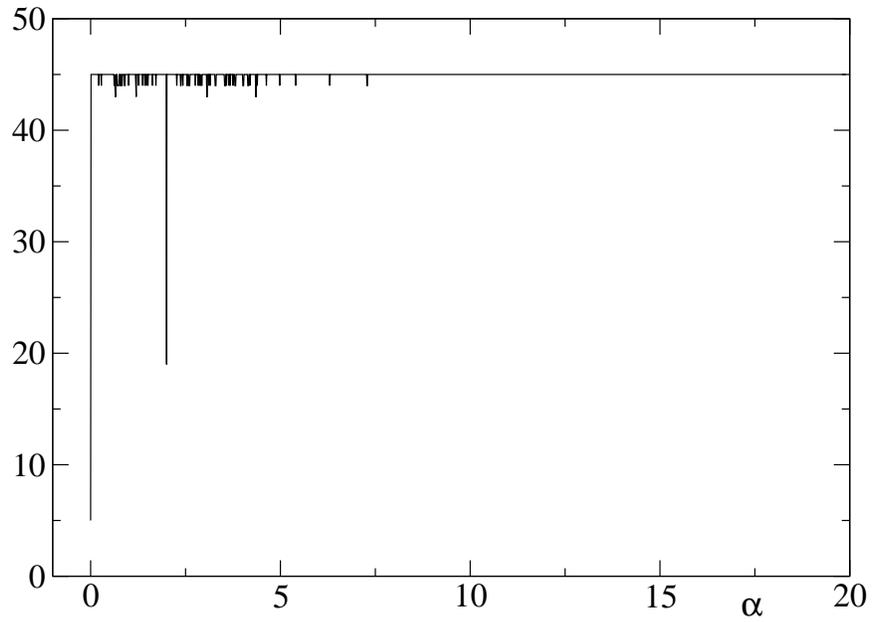}
\end{center}
\caption{\label{sizeofspectrum}Size of the spectrum of $H_8( \alpha )$; that is: number of eigen-energies vs. $\alpha$. No crossings occurr beyond $\alpha \approx 7.29$. The nearest-neighbor interaction model corresponding to $\alpha = \infty$ has $40$ eigen-energies.}
\end{figure}

$\;$
\newpage
$\:$

\begin{figure}
\begin{center}
\psfig{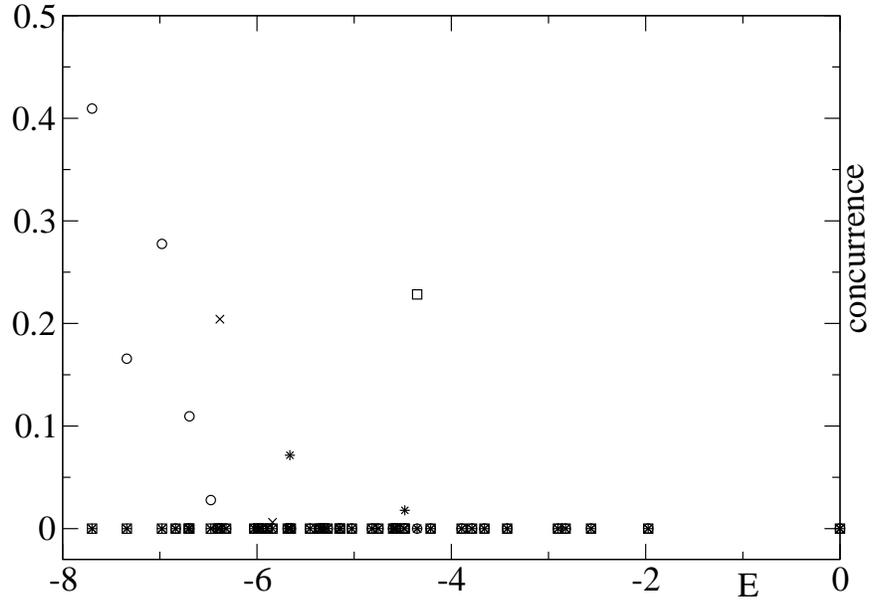}
\end{center}
\caption{\label{a=1} Concurrence of $\rho_{j,k} (E)$ for $N=8$ and $\alpha=1$. $\circ$ corresponds to $|j-k|=1$; $\times$ to $|j-k|=2$; $\ast$ to $|j-k|=3$; and $\Box$ to $|j-k|=4$.}
\end{figure}

\newpage
$\;$

\begin{figure}
\begin{center}
\psfig{figure=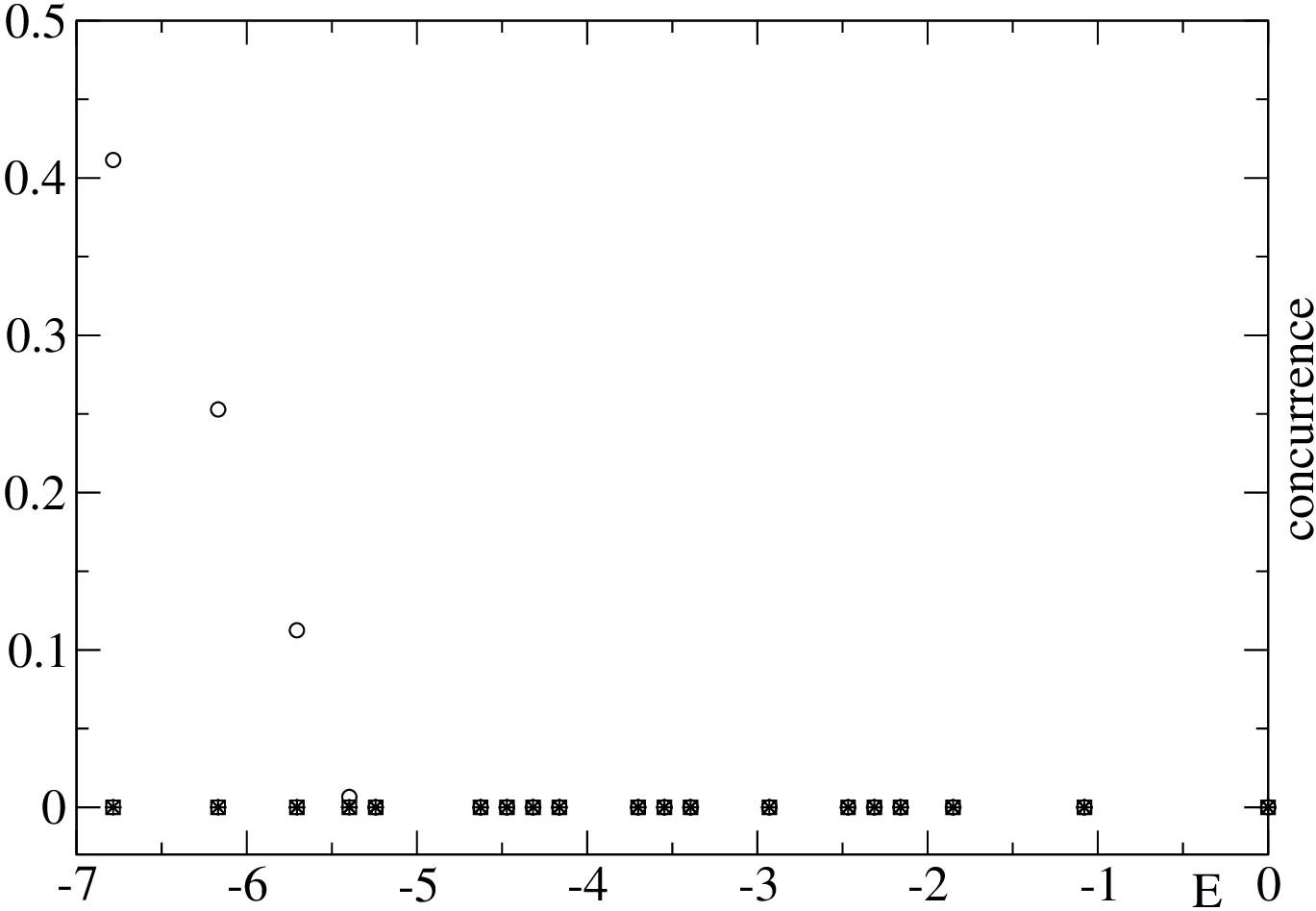,width=11.4cm}
\end{center}
\caption{\label{a=2}Concurrence of $\rho_{j,k} (E)$ for $N=8$ and $\alpha=2$. $|j-k|= 1(\circ),2(\times), 3(\ast), 4(\Box)$.}
\end{figure}

$\;$

\newpage
$\;$

\begin{figure}
\begin{center}
\psfig{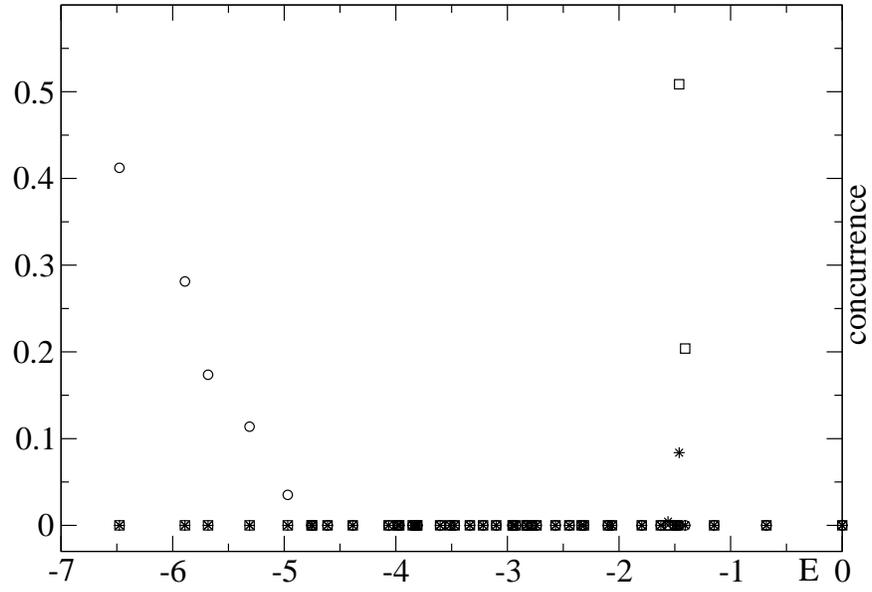}
\end{center}
\caption{\label{a=3}Concurrence of $\rho_{j,k} (E)$ for $N=8$ and $\alpha=3$. $|j-k|= 1(\circ),2(\times), 3(\ast), 4(\Box)$.}
\end{figure}

$\;$

\newpage

$\;$

\begin{figure}
\begin{center}
\psfig{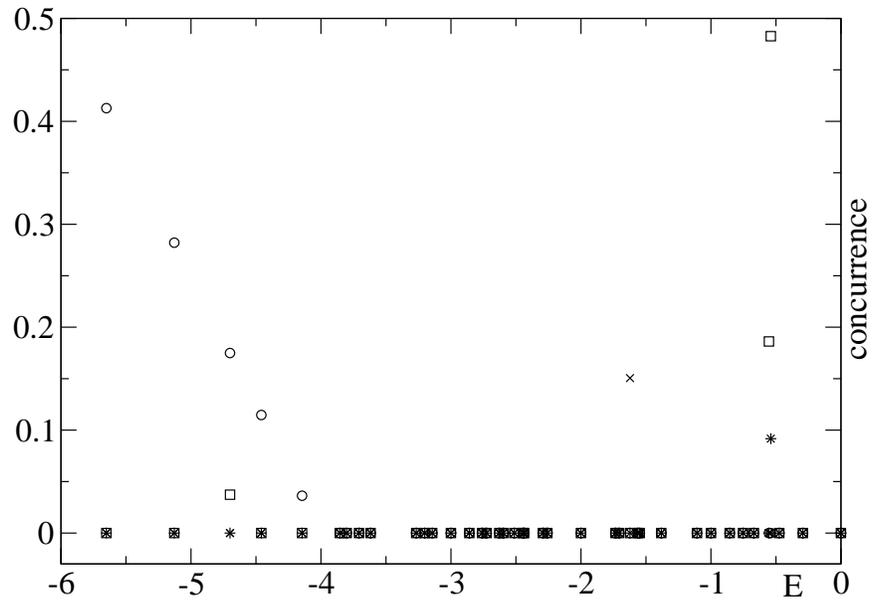}
\end{center}
\caption{\label{a=nn} Concurrence of $\rho_{j,k} (E)$ for $N=8$ in the nearest-neighbor model. $|j-k|= 1(\circ),2(\times), 3(\ast), 4(\Box)$.}
\end{figure}

$\;$

\newpage

$\;$

\begin{figure}
\begin{center}
\psfig{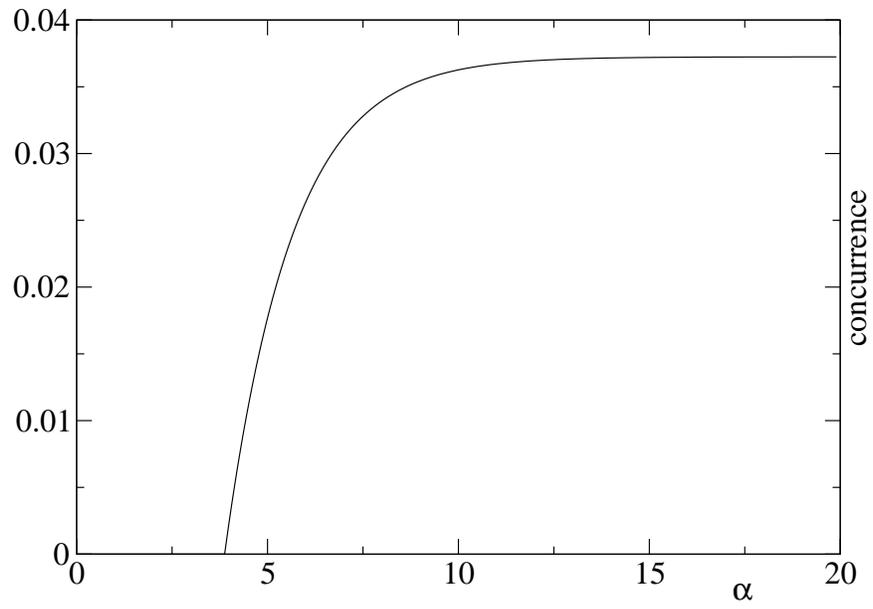}
\end{center}
\caption{\label{15-2}Concurrence of $\rho_{1,5} (E)$ vs. $\alpha$, for $E$ the second excited state for $N=8$.}
\end{figure}

$\;$

\newpage 
$\;$

\begin{figure}
\begin{center}
\psfig{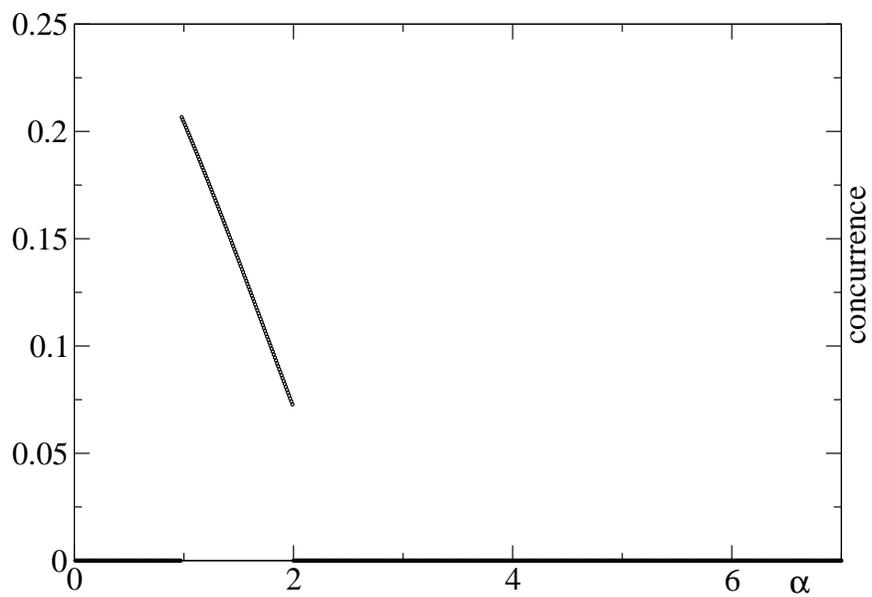}
\end{center}
\caption{\label{14-8}Concurrence of $\rho_{1,3} (E)$ vs. $\alpha$ for $E$ the eighth excited state for $N=8$. }
\end{figure}

$\;$

\newpage

$\;$

\begin{figure}
\begin{center}
\psfig{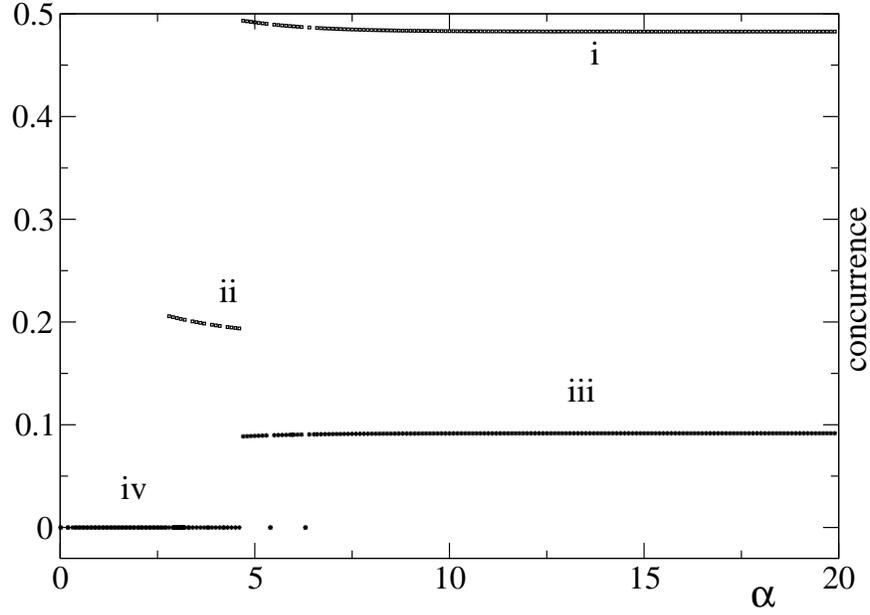}
\end{center}
\caption{\label{15&14-42}Concurrence of $\rho_{1,5}(E)$ ({\em i}, {\em ii}, and part of {\em iv}) and $\rho_{1,4}(E)$ ({\em iii}, and part of {\em iv}) vs. $\alpha$ for the fourty first excited state for $N=8$. Both functions show discontinuities below $\alpha \approx 7.29$. In the curve labeled {\em iv} corresponding to zero concurrence, points for distance $|j-k|=4$ are interspersed with points for distance $| j-k|=3$. See figure \ref{detail}.}
\end{figure}

$\;$

\newpage

$\;$

\begin{figure}
\begin{center}
\psfig{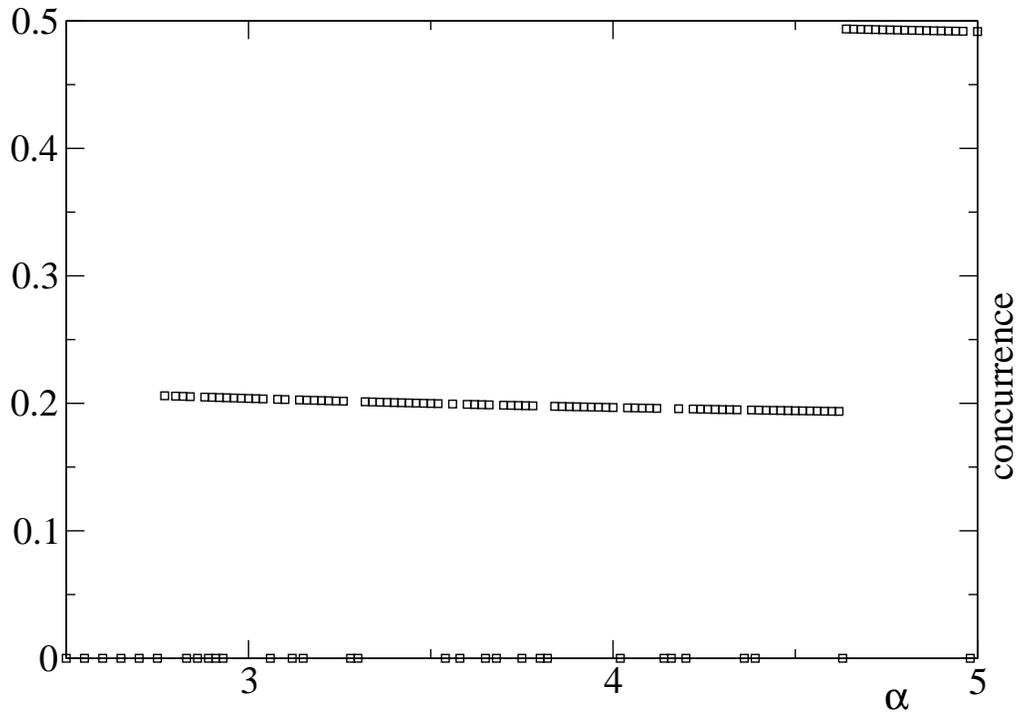}
\end{center}
\caption{\label{detail}Detail of the concurrence  of $\rho_{1,5}(E)$  vs. $\alpha$ for the fourty first excited state for $N=8$. From top to bottom the curves correspond to those labeled {\em i}, {\em ii} and {\em iv} in the previous figure \ref{15&14-42}.}
\end{figure}

$\;$

\newpage

$\;$

\begin{figure}
\begin{center}
\psfig{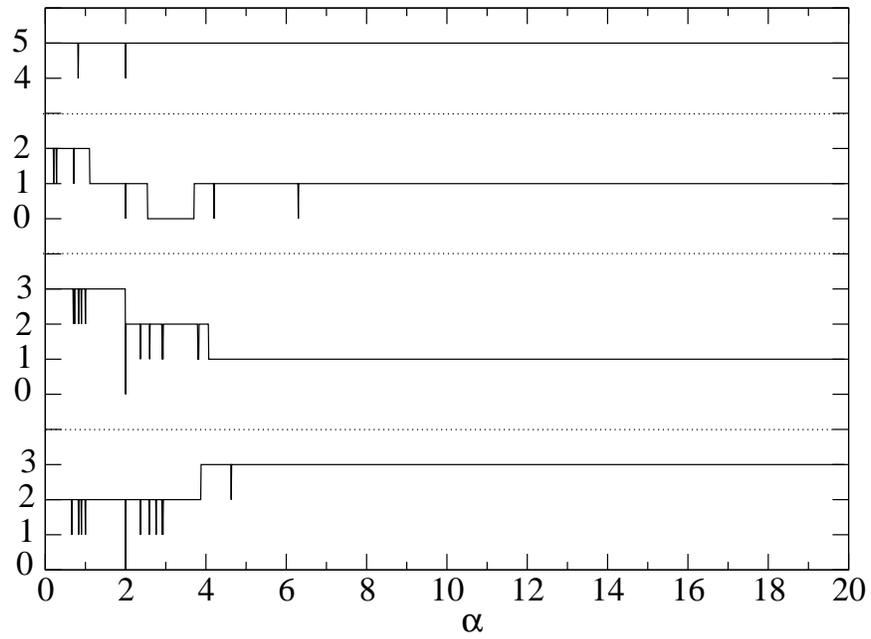}
\end{center}
\caption{\label{cantidad}Number of entangled energy eigen-states for the four possible distances for $N=8$ as a function of $\alpha >0$. From top to bottom, $|j-k|=1,2,3,4$. }
\end{figure}

$\;$

\newpage
$\;$

\begin{figure}
\begin{center}
\psfig{figure=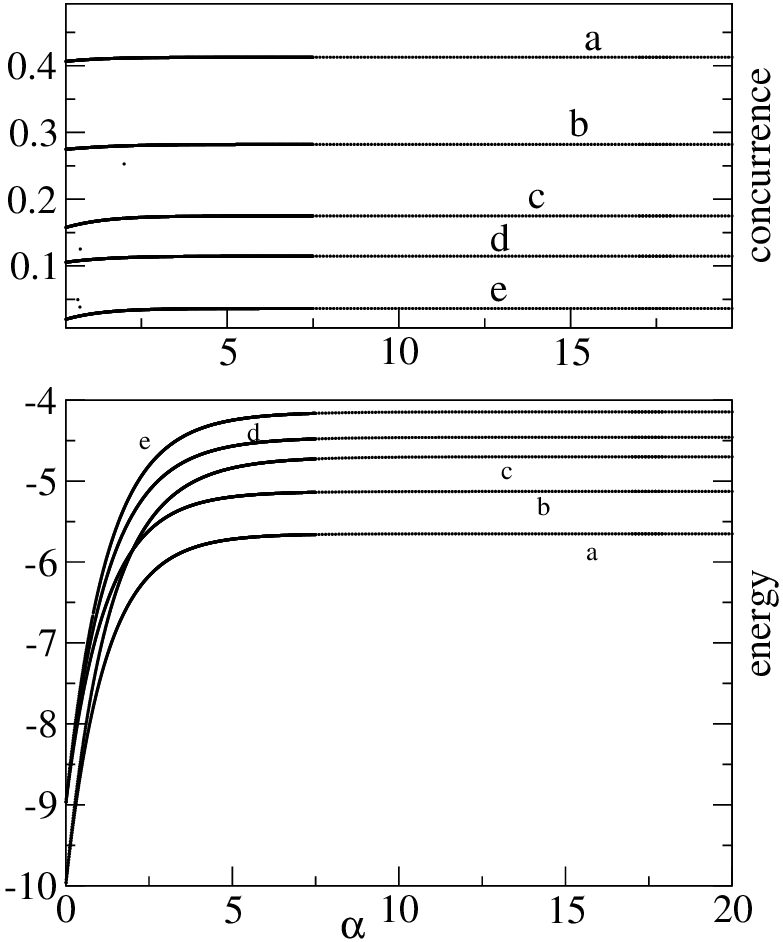,width=11.4cm}
\end{center}
\caption{\label{fig12-13}Top graph: Nearest neighbor concurrence vs $\alpha$ for $N=8$. The top curve labeled {\em a} corresponds to the ground-state.  Not all discontinuities are visible in the figure.\\
Lower graph: The corresponding  five energies vs $\alpha$  curves which carry nearest neighbor entanglement for $N=8$. The curve labels {\em a-e} correspond to the concurrence curves of the same label in the top graph. The dimension of the associated orthoprojectors are: 1 (a), 3 (b), 1 (c), 6 (d) and 6 (e).}
\end{figure}

$\;$

\newpage
$\;$

\begin{figure}
\begin{center}
\psfig{figure=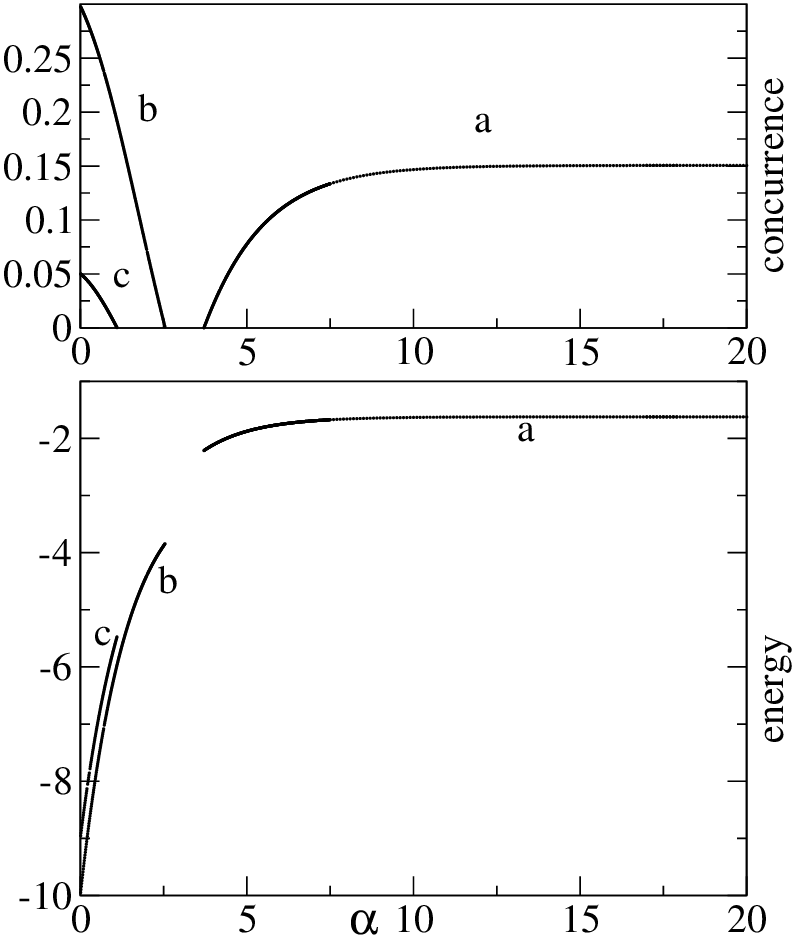,width=11.4cm}
\end{center}
\caption{\label{fig14-15}Top graph: Next-nearest neighbor concurrence vs $\alpha$ for $N=8$.\\
Lower graph: The three energies vs $\alpha$  curves which carry next-nearest neighbor entanglement for $N=8$. The labels on the curves correspond to those on the concurrence curves of the top graph. The dimensions of the associated orthoprojectors are: 1 (a), 1 (b), and 6 (c).}
\end{figure}

$\;$

\newpage

$\;$

\begin{figure}
\begin{center}
\psfig{figure=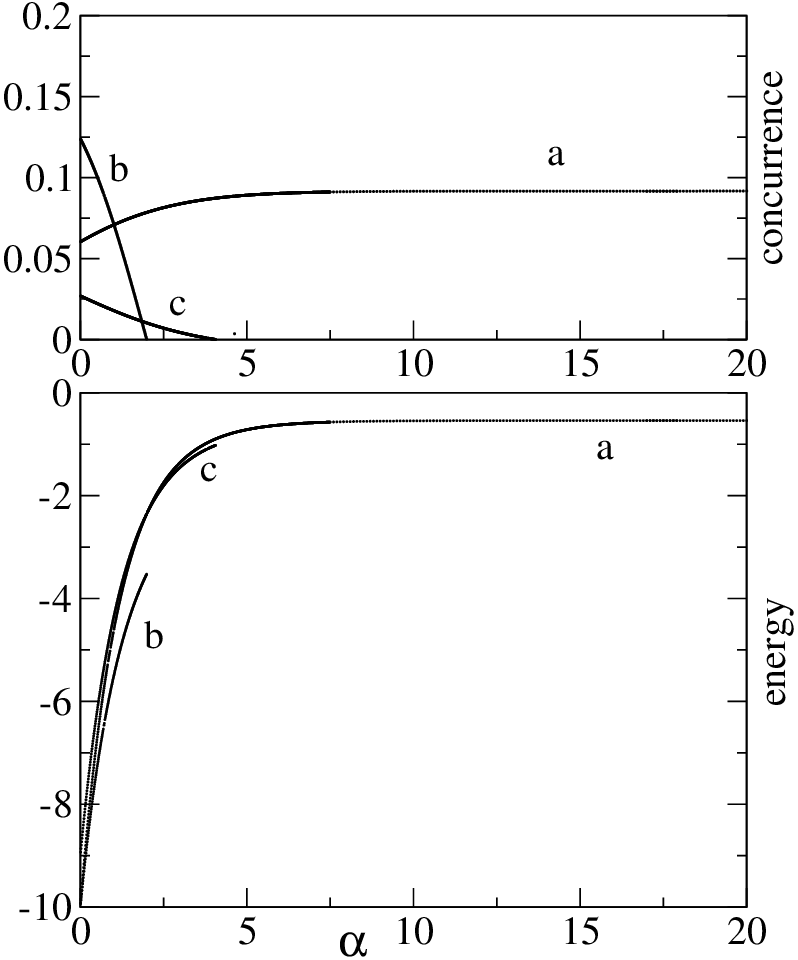,width=11.4cm}
\end{center}
\caption{\label{fig16-17}Top graph: Concurrence vs $\alpha$ for $N=8$ and $|j-k|=3$.\\
Lower graph: The three energies vs $\alpha$  curves which carry entanglement for $N=8$ at distance $|j-k|=3$. The curve labels correspond to those of the top graph. Curves {\em a} and {\em c} cross at $\alpha \approx 2.35$. The dimensions of the associated orthoprojectors are: 1 (a), 1 (b), and 3 (c).}
\end{figure}

$\;$

\newpage

$\;$

\begin{figure}
\begin{center}
\psfig{figure=fig18-19.eps,width=11.4cm}
\end{center}
\caption{\label{fig18-19}Top graph: Concurrence vs $\alpha$ for $N=8$ and maximal distance $|j-k|=4$.\\
Lower graph: The three energies vs $\alpha$  curves which carry entanglement for $N=8$ at maximal distance. The curve labels correspond to those of the top graph. Curves {\em a} and {\em b} cross at $\alpha \approx 4.63$ where the concurrence curves {\em a} and {\em b} have a discontinuity. The dimensions of the associated orthoprojectors are: 1 (a), 6 (b), and 1 (c).}
\end{figure}

$\;$

\newpage

$\;$
\begin{figure}
\begin{center}
\psfig{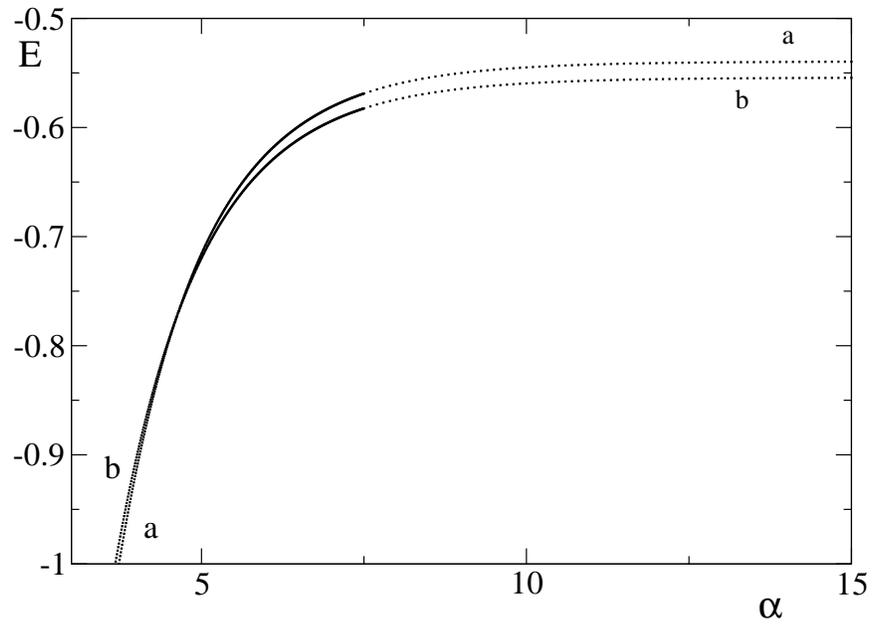}
\end{center}
\caption{\label{ccrE_15a}Detail of the crossing of the energy curves {\em a} and {\em b} in the lower graph of the previous figure \ref{fig18-19}. The curves do not coincide to the right of the crossing, a feature  which is not resolved at the scale of figure \ref{fig18-19}.}
\end{figure}

\end{document}